# T-Base: A Triangle-Based Iterative Algorithm for Smoothing Quadrilateral Meshes

Gang Mei[1], John C.Tipper[1] and Nengxiong Xu[2]

**Abstract** We present a novel approach named T-Base for smoothing planar and surface quadrilateral meshes. Our motivation is that the best shape of quadrilateral element – square – can be virtually divided into a pair of equilateral right triangles by any of its diagonals. When move a node to smooth a quadrilateral, it is optimal to make a pair of triangles divided by a diagonal be equilateral right triangles separately. The finally smoothed position is obtained by weighting all individual optimal positions. Three variants are produced according to the determination of weights. Tests by the T-Base are given and compared with Laplacian smoothing: The Vari.1 of T-Base is effectively identical to Laplacian smoothing for planar quad meshes, while Vari.2 is the best. For the quad mesh on underlying parametric surface and interpolation surface, Vari.2 and Vari.1 are best, respectively.

## 1 Introduction

The quality of meshes is critical to obtain reliable simulation results in finite element analyses. Usually after generating computational meshes, it is necessary to improve the quality of meshes in further. There are two important categories of quality improvement methods. One is called *clear-up* techniques, which alters the connectivity between elements. The other is called *mesh smoothing*, which only relocates the nodes. There are numerous publications on the topic of mesh smoothing. And we just refer some popular and representative ones.

Gang Mei, John C.Tipper (✉)
Institut für Geowissenschaften – Geologie, Albert-Ludwigs-Universität Freiburg, Albertstr. 23B, D-79104, Freiburg im Breisgau, Germany
e-mail: {gang.mei, john.tipper}@geologie.uni-freiburg.de

Nengxiong Xu (✉)
School of Engineering and Technology, China University of Geosciences, Beijing,100083, China
e-mail: xunengxiong@yahoo.com.cn

The most popular smoothing methods is Laplacian smoothing [5, 7], which repositions each node at the centroid of its neighbouring nodes in one iteration. The popularity of this method comes from its efficiency and effectiveness. A simpler but more effective method is angle-based approach [16], in which new locations are calculated by conforming specific angle ratios in surrounding polygons.

A geometric element transformation method [14], which is based on a simple geometric transformation, is proposed and applied to polygons. Shimada et al [12] proposed a method which treats nodes as the centre of bubbles and nodal locations are obtained by deforming bubbles with each other.

A projecting/smoothing method is proposed for smoothing surface meshes [4], where the new position of each free node is obtained by minimizing the mean ratio of all triangles sharing the free node. Based on quadric surface fitting and by combining vertex projecting, curvature estimating and mesh labelling, Wang and Yu [15] proposed a novel method and applied it in biomedical modelling.

A variational method for smoothing surface and volume triangulations is proposed by Jiao [9], where the discrepancies between actual and target elements is reduced by minimizing two energy functions. Also, a general-purpose algorithm called the target-matrix paradigm is introduced in [10], and can be applied to a wide variety of mesh and element types.

To smooth meshes better, two or more basic methods can be combined into a hybrid approach [1,2,6], i.e., an analytical framework for mesh quality metrics and optimization direction computation in physical and parametric space are proposed for smoothing surface quad meshes in [13].

In this paper we introduce a novel iterative method named T-Base to smooth planar and surface quad meshes. The best shape of a quadrilateral element is square, which can be virtually divided into a pair of equilateral right triangles by any of its diagonals. Hence, when move a node to smooth a quad element, it is optimal to make the two triangles divided the diagonal consisted by the node and its opposite one be equilateral right triangles separately. The final smoothed position is obtained by weighting all the separate optimal positions.

When smooth surface quad meshes, we firstly compute the local coordinates system for each virtual triangle and then calculate the optimal position, and finally obtain the smoothed node by transforming it from local coordinates to the global coordinates and weighting all individual optimal positions.

After generating the optimal smoothed positions, they should be moved again in order to preserve the features of initial surfaces. For quad mesh on parametric surfaces, we project the smoothed node onto the original parametric surface along the normal. For quad mesh on interpolation surfaces, we re-interpolate the smoothed nodes to fit them with the initial surfaces.

The rest of this paper is organized as follows. Sect.2 describes the details of the T-Base including its three variants for smoothing planar quad meshes. In Sect.3, we simply extend T-Base to smooth surface quad meshes. Then we give several examples in Sect.4 to test the performance of the T-Base and compare it with Laplacian smoothing. Finally, Sect.5 concludes this work.

## 2 T-Base for Planar Quad Meshes

The best shape of a quadrilateral element is square, which can be virtually divided into a pair of equilateral right triangles by any of its diagonals. When move a node in order to smooth a quadrilateral, it is optimal to make the two triangles divided by the diagonal consisted with the node and its opposite one be equilateral right triangles separately (Fig.1).

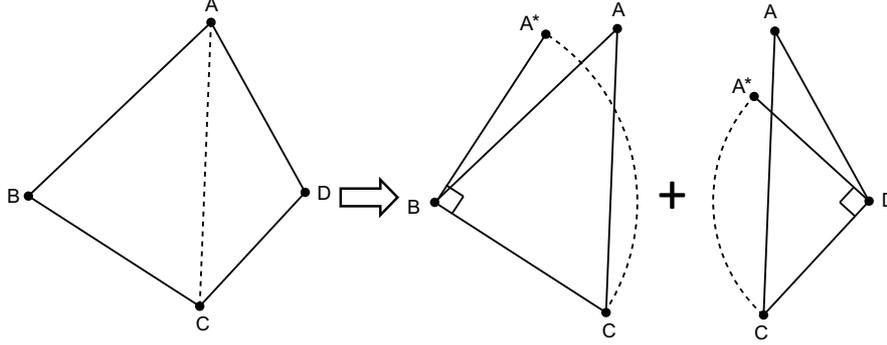

**Fig. 1** Smoothing of quad element ABCD based on a pair of virtual triangles

Consider a single quadrilateral element ABCD shown in Fig.1. It is virtually divided into a pair of triangles ABC and CDA. $A^*$s are the positions to which node A would have to be moved to make ABC and CDA be equilateral right triangles, assuming that nodes B, C and D were fixed. The coordinates of $A^*$ in triangles ABC and CDA are:

$$\begin{cases} X_A^* = X_B + Y_B - Y_C \\ Y_A^* = -X_B + Y_B + X_C \end{cases},\; \begin{cases} X_A^* = X_D - Y_D + Y_C \\ Y_A^* = X_D + Y_D - X_C \end{cases} \quad (1)$$

Now assume ABCD is part of a quadrilateral mesh. Each node of ABCD –for instance node A – is then shared with several other elements, and $A^*$ can be calculated for each of these. The final position of A – its optimal smoothed position – is obtained by considering all the separately calculated $A^*$'s (Fig.2).

The traditional Laplacian smoothing assumes that the new position of a node should be an average of the positions calculated for it for each of the elements of which it is part (Fig.2). This assumption is not essential, however, and it can be relaxed by making that new position a weighted average of those positions, with the weights being proportional to the lengths of the opposite edges. Then, for node A:

$$X_A = \sum_{i=1}^{2n}(w_i \cdot X_{Ai}^*),\; Y_A = \sum_{i=1}^{2n}(w_i \cdot Y_{Ai}^*) \quad (2)$$

, where $n$ is the number of elements of which node A is part, and $w_i$ is the weight of the $i^{th}$ separate position $A^*$ which can be calculated according to the length $l_i$ of relevant $i^{th}$ edge.

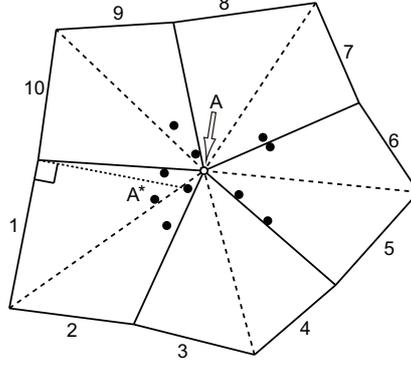

**Fig. 2** Node A belongs to 5 quad elements. $A^*$ can be calculated for each triangle separately (black circles). Optimal smoothed position for A is produced from all $A^*$s

According to the determination of the weights $w_i$, we produce three variants:
**Variant 1**. This variant is termed as the *average* version of the T-Base:

$$w_i = \frac{1}{2n} = l_i^0 \Big/ \sum_{i=1}^{2n} l_i^0 \qquad (3)$$

The averaging process is effectively identical to that used in the traditional Laplacian smoothing, which explains why test results obtained using Laplacian smoothing are identical to that of T-Base for planar quad meshes (see Figs 3, 4). But the above conclusion is no longer true for surface quadrilateral meshes.

**Variant 2**. This variant is termed as the (−1/2) *inverse-length* version:

$$w_i = \frac{1}{\sqrt{l_i}} \Big/ \sum_{i=1}^{2n} \frac{1}{\sqrt{l_i}} = l_i^{-1/2} \Big/ \sum_{i=1}^{2n} l_i^{-1/2} \qquad (4)$$

**Variant 3**. This variant is termed as the (−1) *inverse-length* version:

$$w_i = \frac{1}{l_i} \Big/ \sum_{i=1}^{2n} \frac{1}{l_i} = l_i^{-1} \Big/ \sum_{i=1}^{2n} l_i^{-1} \qquad (5)$$

The introduction of inverse-length weighting is in many respects advantageous, because high quality elements such as equilateral quadrilateral element or even square generally have nearly or exactly same-length edges. In order to transform quadrilateral elements to be equilateral as more as possible, we let longer edges of an element have less importance in the smoothing than shorter ones.

The disadvantage of inverse-length versions (**Vari.2** and **3**) over the average version (**Vari.1**) is of course that it brings a time penalty, as the weights have to be calculated afresh at each iteration step. This is also the reason why inverse-length versions need more iteration steps to converge than that of the average version.

The implementation of T-Base for planar quadrilateral meshes is very simple: (1) search all incident elements for each node; (2) calculate of smoothed positions of each node by making relevant virtual triangles be equilateral right; (3) iterate previous step until a tolerance distance is reached.

## 3 T-Base for Surface Quad Meshes

Eq.1 computes the optimal position for a virtual triangle in 2D. For surface quad meshes, we firstly compute the local coordinates system for each virtual triangle and then calculate the optimal position via Eq.1, and finally obtain the smoothed nodal position by recovering it to global coordinates and weighting all $A^*$s.

After obtaining the optimal smoothed positions, updating should be done for different type of discrete surfaces in order to preserve the shape of initial surfaces. For quad mesh on parametric surfaces, we compute the normal at each vertex and then project the smoothed node onto the original parametric surface along the normal to obtain final position (Fig.5). For quad mesh on interpolation surfaces, we re-interpolate the smoothed nodes to fit them with the initial surfaces (Fig.6).

Flow of the T-Base for surface quad meshes is listed in Algorithm.1.

**Algorithm 1** T-Base for Smoothing Surface Quad Meshes

*Input*: An original surface quad mesh
*Output*: A smoothed surface quad mesh
1: Search the incident elements for each node $v_i$.
2: **while** iterations not terminate **do**
3:    **for** each node $v_i$ **do**
4:       **for** each incident element $Q_j$ ($0 \leq j < n$) of $v_i$ **do**
5:          Divide $Q_j$ into two triangles and calculate local coordinates separately.
6:          Obtain a pair of $A^*$'s locally and transform them back to global.
7:          Calculate a pair of weights $w$'s in $Q_j$.
8:       **end for**
9:       Obtain optimal smoothed position of $v_i$:
$$X_A = \sum_{i=1}^{2n}(w_i \cdot X_{Ai}^*), Y_A = \sum_{i=1}^{2n}(w_i \cdot Y_{Ai}^*), Z_A = \sum_{i=1}^{2n}(w_i \cdot Z_{Ai}^*)$$
10:      Update $v_i$ by projecting it to initial parametric surface or re-interpolating.
11:    **end for**
12: **end while**

## 4 Applications and Discussion

### *4.1 Mesh Quality*

The simplest way to measure mesh quality is to calculate the distortion values for each of the mesh elements separately, and then to compare the distributions, including mean quality (**MQ**) and mean square error (**MSE**), of those values. For a quad ABCD, we use the measure $\lambda$ [8], shown in Eq.6. The value of $\lambda$ lies between 0 and 1; $\lambda = 0$ when any three nodes are collinear; $\lambda = 1$ when ABCD is square.

$$\lambda = 2\sqrt[4]{\frac{\|AB \times AD\| \cdot \|BC \times BA\| \cdot \|CD \times CB\| \cdot \|DA \times DC\|}{(\|AB\|^2 + \|AD\|^2)(\|BC\|^2 + \|BA\|^2)(\|CD\|^2 + \|CB\|^2)(\|DA\|^2 + \|DC\|^2)}} \quad (6)$$

For a quad element in 3D, it's warped generally. In this paper, we propose a new measurement $\gamma$ in which shape and warpage are taken into account. A quad ABCD can be divided into four triangles: ABC, BCD, CDA and DAB. We firstly calculate the local coordinates system of these triangles and project the original quad element ABCD onto each local coordinates system to obtain four planar quads $ABCD^P$, $BCDA^P$, $CDAB^P$ and $DABC^P$, respectively. Let $\lambda_1$, $\lambda_2$, $\lambda_3$ and $\lambda_4$ denote the $\lambda$ values of the four planar quads, then $\gamma = (\lambda_1 + \lambda_2 + \lambda_3 + \lambda_4) / 4$. The value of $\gamma$ also lies between 0 and 1; we specially set $\lambda = 0$ when any three nodes are collinear; $\lambda = 1$ when ABCD is coplanar square. Thus, we have:

$$MQ = \frac{1}{n}\sum_{i=1}^{n} \gamma_i, \quad MSE = \sqrt{\frac{1}{n}\sum_{i=1}^{n}(\gamma_i - MQ)^2} \quad (7)$$

, where *n* is the number of elements in a quad mesh.

### *4.2 Tests of Smoothing Planar Quad Meshes*

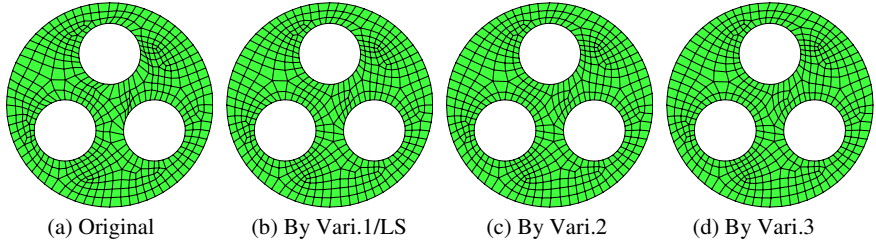

(a) Original  (b) By Vari.1/LS  (c) By Vari.2  (d) By Vari.3

**Fig. 3** Test 1 of smoothing planar quad mesh by T-Base and Laplacian smoothing (LS)

These two original quad meshes are generated by Q-Morph [11]. Fig.3 and Fig.4 display the results of the two planar quad meshes by Laplacian smoothing and T-Base. From the comparison of mesh quality listed in Table.1, we can learn that Vari.1 is effectively identical to Laplace smoothing, Vari.3 and Vari.2 are better than Vari.1, but Vari.2 is best.

**Convergence** We do not give the algebraic proof of convergence for T-Base in theory. But our tests show that T-Base does converge for planar meshes, and the numbers of iteration steps of Vari.1, 2 and 3 always increases when it converges.

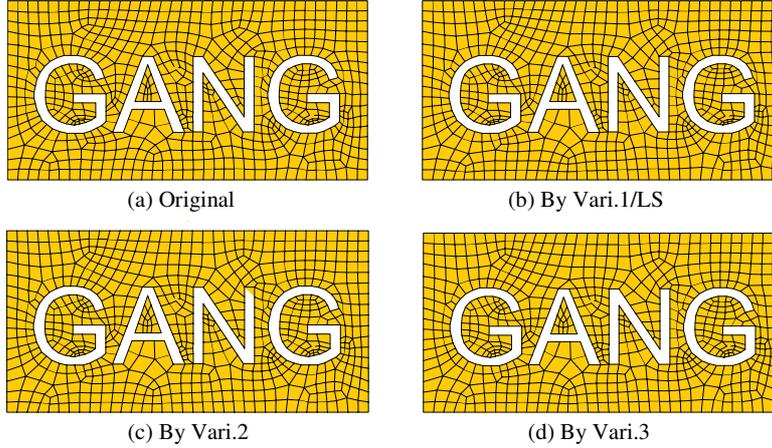

(a) Original  (b) By Vari.1/LS

(c) By Vari.2  (d) By Vari.3

**Fig. 4** Test 2 of smoothing planar quad mesh by T-Base and Laplacian smoothing (LS)

## 4.3 Tests of Smoothing Surface Quad Meshes

We firstly generate the planar meshes in a circular area, and then project it onto the parametric surface $z = 200 - 0.02(x^2 + y^2)$ to obtain the mesh in Fig.5a. The original mesh in Fig.6a is for height interpolation by Kriging method [3]; only $z$-value/height is interpolated while coordinates $x$ and $y$ are fixed.

Fig.5 shows the results of quad meshes on a underlying parametric surface. Noticeably, only Laplacian smoothing converges after 49 iterations. We just let T-Base iterate 49 times as that of Laplacian smoothing. From the comparison of mesh quality in Table.1, we can learn that T-Base is better than Laplacian smoothing. In further, Vari.2 is the best, and the Vari.1 is better than Vari.3 since the distribution of element qualities is better than that of Vari.3.

In Fig.6, all optimal smoothed by only Laplacian smoothing or T-Base is generated firstly, and then $z$-value is re-interpolated by Kriging method. Due to the expensive cost of re-interpolation, we only iterate 10 times. Similar to the quad mesh on parametric surface, T-Base is better than Laplacian smoothing. But Vari.1 is the best, and then the Vari.2, while Vari.3 is the worst.

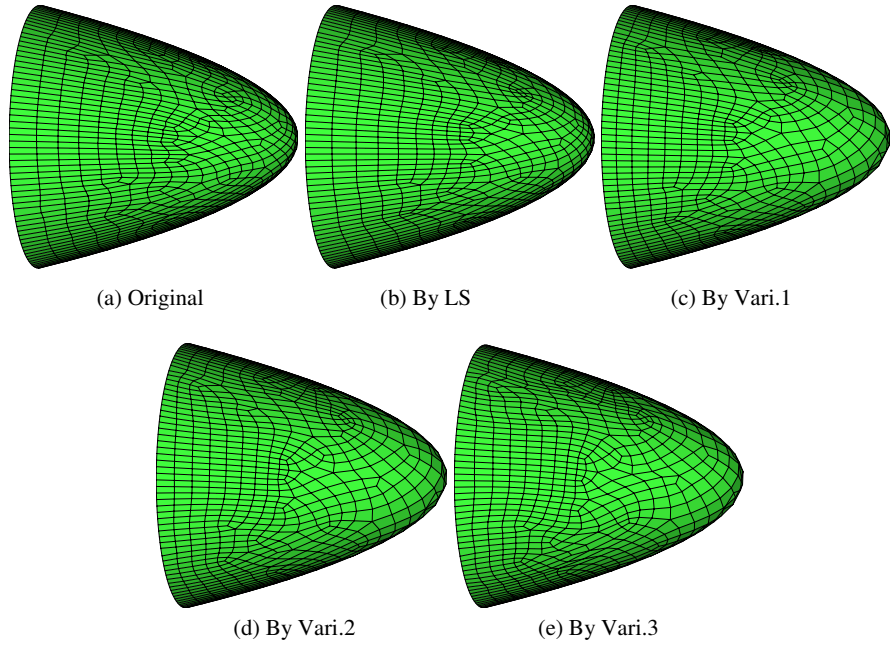

**Fig. 5** Smoothing results of surface quad mesh on underlying parametric surface

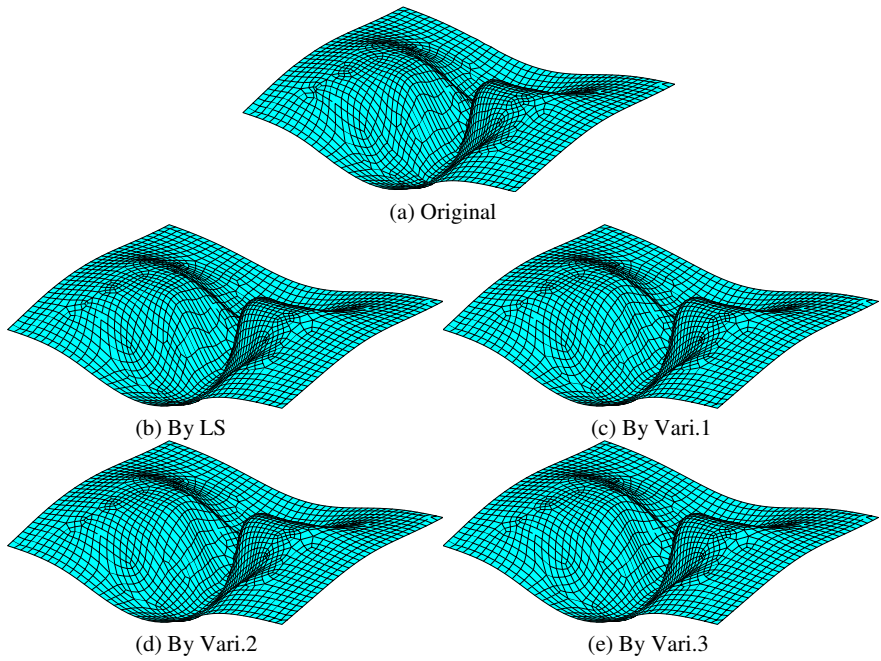

**Fig. 6** Smoothing results of surface quad mesh on interpolation surface

**Convergence** Only Laplacian smoothing converges for the quad meshes on parametric underlying surface. According to the convergence analysis for planar quad meshes, T-Base needs to iterate more times than Laplacian smoothing, hence, we can firstly record the iteration number of Laplacian smoothing for surface quad meshes, and then set the number from Laplacian smoothing to be the maximum iterations in T-Base. This is the reason we only iterate 49 times in T-Base. When smooth quad meshes on interpolation surface, since re-interpolation is too expensive, we just test the results after a specified-number of iterations. This solution of ending iterations is acceptable and valuable in practical applications.

**Table 1** Mesh quality results of smoothing planar and surface quad meshes

| Mesh | Algorithm | Element quality (0.0~1.0) | | | | | MQ | MSE |
|---|---|---|---|---|---|---|---|---|
| | | 0.0~0.2 | 0.2~0.4 | 0.4~0.6 | 0.6~0.8 | 0.8~1.0 | | |
| Fig.3 | Original | 0.00% | 0.00% | 1.13% | 6.53% | 92.34% | 0.9327 | 0.0860 |
| | Vari.1/LS | 0.00% | 0.00% | 0.00% | 2.93% | 97.07% | 0.9514 | 0.0543 |
| | Vari.2 | 0.00% | 0.00% | 0.00% | 3.15% | 96.85% | 0.9528 | 0.0520 |
| | Vari.3 | 0.00% | 0.00% | 0.00% | 3.60% | 96.40% | 0.9525 | 0.0535 |
| Fig.4 | Original | 0.00% | 0.00% | 0.45% | 6.47% | 93.08% | 0.9277 | 0.0781 |
| | Vari.1/LS | 0.00% | 0.00% | 0.30% | 6.47% | 93.23% | 0.9348 | 0.0764 |
| | Vari.2 | 0.00% | 0.00% | 0.45% | 5.86% | 93.69% | 0.9369 | 0.0722 |
| | Vari.3 | 0.00% | 0.00% | 0.90% | 4.81% | 94.29% | 0.9357 | 0.0753 |
| Fig.5 | Original | 0.00% | 0.99% | 67.26% | 18.81% | 12.94% | 0.5916 | 0.1457 |
| | LS | 0.00% | 0.00% | 62.45% | 26.03% | 11.53% | 0.6021 | 0.1380 |
| | Vari.1 | 0.00% | 0.00% | 0.64% | 75.74% | 23.62% | 0.7392 | 0.1020 |
| | Vari.2 | 0.00% | 0.00% | 3.18% | 61.10% | 35.71% | 0.7565 | 0.1107 |
| | Vari.3 | 0.14% | 1.13% | 10.04% | 48.30% | 40.38% | 0.7558 | 0.1356 |
| Fig.6 | Original | 0.00% | 0.00% | 2.40% | 9.47% | 88.12% | 0.9191 | 0.1020 |
| | LS | 0.00% | 0.00% | 1.98% | 8.01% | 90.01% | 0.9309 | 0.0959 |
| | Vari.1 | 0.00% | 0.00% | 0.14% | 5.70% | 94.16% | 0.945 | 0.0727 |
| | Vari.2 | 0.00% | 0.00% | 0.99% | 5.04% | 93.97% | 0.9447 | 0.0803 |
| | Vari.3 | 0.00% | 0.28% | 1.37% | 4.90% | 93.45% | 0.9434 | 0.0902 |

## 5 Conclusion

We present a novel iterative smoothing algorithm called T-Base for planar and surface quad meshes based on virtually dividing a quad element into a pair of triangles by its diagonal. We relocate a node by making all of the incident virtual triangles be equilateral right triangles separately, and then weighting all separate smoothed positions. According to the determination of weights based on length of relevant edges, three variants of T-Base are produced. The T-Base is applied on planar and surface quad meshes, and compared with Laplacian smoothing. The Vari.1 of T-Base is effectively identical to Laplacian smoothing for planar quad meshes, while Vari.2 and 3 are better. For the quad mesh on a underlying parametric surface, Vari.2 is the best; and Vari.1 is the best for the quad mesh on a interpolation surface. Tests also show that T-Base always converges for planar meshes.

**Acknowledgments**. This research was supported by the Natural Science Foundation of China (Grant Numbers 40602037 and 40872183) and the Fundamental Research Funds for the Central Universities of China.